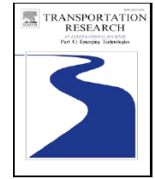

# Short-term origin-destination demand prediction in urban rail transit systems: A channel-wise attentive split-convolutional neural network method

Jinlei Zhang[a], Hongshu Che[c], Feng Chen[b, d, *], Wei Ma[e], Zhengbing He[f]

[a] State Key Laboratory of Rail Traffic Control and Safety, Beijing Jiaotong University, Beijing 100044, China
[b] School of Civil Engineering, Beijing Jiaotong University, Beijing 100044, China
[c] School of Automation, Southeast University, Nanjing, 211189, China
[d] Beijing General Municipal Engineering Design and Research Institute Company Ltd., Beijing 100082, China
[e] Department of Civil and Environmental Engineering, The Hong Kong Polytechnic University, Hong Kong SAR, China
[f] Beijing Key Laboratory of Traffic Engineering, Beijing University of Technology, Beijing 100124, China
* Correspondence: fengchen@bjtu.edu.cn

**Abstract**

Short-term origin-destination (OD) flow prediction in urban rail transit (URT) plays a crucial role in smart and real-time URT operation and management. Different from other short-term traffic forecasting methods, the short-term OD flow prediction possesses three unique characteristics: 1) data availability: real-time OD flow is not available during the prediction; 2) data dimensionality: the dimension of the OD flow is much higher than the cardinality of transportation networks; 3) data sparsity: URT OD flow is spatiotemporally sparse. There is a great need to develop novel OD flow forecasting method that explicitly considers the unique characteristics of the URT system. To this end, a channel-wise attentive split–convolutional neural network (CAS-CNN) is proposed. The proposed model consists of many novel components such as the channel-wise attention mechanism and split CNN. In particular, an inflow/outflow-gated mechanism is innovatively introduced to address the data availability issue. We further originally propose a masked loss function to solve the data dimensionality and data sparsity issues. The model interpretability is also discussed in detail. The CAS–CNN model is tested on two large-scale real-world datasets from Beijing Subway, and it outperforms the rest of benchmarking methods. The proposed model contributes to the development of short-term OD flow prediction, and it also lays the foundations of real-time URT operation and management.

*Keywords:* Deep learning; Urban rail transit; Short-term origin-destination prediction; Channel-wise attention; Split CNN

# 1 Introduction

In recent years, the urban rail transit (URT) has experienced rapid expansion. Significant attention has been devoted to its intelligent operation and management. As one of the fundamental tasks of intelligent transportation systems, short-term passenger flow prediction has attracted increasing research interest because of its practical influence on both passengers and operators (Liu *et al.*, 2020). For operators, the result of OD prediction can help to better monitor the real-time spatiotemporal distribution of passenger flows, thus supporting decisions on network management tasks, such as implement congestion control and anomaly detection. Real-time measures, such as the adjustment of train timetables (shortening or extending the headway), can be taken to avoid congestions and save operational costs. If the congestion or large passenger flow is monitored, operators can reasonably allocate staff members to evacuate passengers, thus avoiding or improving the accident situation. Moreover, for passengers, an accurate short-term prediction is beneficial to route scheduling that saves travel time and thus improves travel experience.

Network-scale short-term passenger flow prediction systems for URT can be categorized into inflow prediction, origin-destination (OD) passenger flow prediction, and sectional passenger flow prediction (Zhang *et al.*, 2019). Short-term inflow prediction, which refers to the forecast of passenger demands entering each station, has been extensively studied (Han *et al.*, 2019, Liu *et al.*, 2019, Wei and Chen, 2012, Zhang *et al.*, 2019, Zhang *et al.*, 2020, Zhang *et al.*, 2020a). After obtaining the real-time inflow, that is, the passenger origins, short-term OD flow prediction can be conducted to forecast passenger destinations (Vlahogianni *et al.*, 2014). Lastly, the sectional passenger flow prediction task refers to the forecast of the specific path chosen by passengers in order to arrive at destinations from origins. Because the individual's trajectory is difficult to obtain in URT, the sectional passenger flow is usually obtained by leveraging transit assignment models to estimate the travelers' behaviors using OD matrices as essential inputs. Overall, OD flow prediction is the bridge between inflow prediction and sectional passenger flow prediction, and it plays a crucial role in the network-scale short-term passenger flow prediction systems. An accurate OD flow prediction model can provide the spatiotemporal mobility patterns among subway stations, thus contributing to a better understanding of travel behaviors (Xiong *et al.*, 2019). Therefore, this paper focuses on the short-term OD passenger flow prediction in URT.

Short-term network-wise traffic prediction has been studied extensively over the past decades. Various data-driven and model-based prediction methods have been developed to forecast road speed/flow, road OD demand, ride-hailing demand, ride-hailing OD demand, and URT inflow/outflow. To the best of our knowledge, there are few studies on the short-term URT OD passenger flow prediction, as there exist several characteristics that distinguish OD prediction tasks in URT from the rest of prediction tasks. These characteristics can be listed as follows:

1. ***Data availability.*** In most of the short-term traffic forecasting problems, the real-time traffic data can be obtained in time. For example, in the traffic speed prediction task, we can obtain the real-time traffic speed at time $t$ in order to predict the speed at time $t+1$. Similarly, the URT inflow/outflow can be obtained in real-time because passengers must swipe cards when they enter subway stations. The card swiping information can be aggregated in real-time and the inflow/outflow at each URT station can be computed. For this type of tasks, the actual traffic states in the last several time intervals can be used as model inputs when short-term traffic forecast is conducted. However, URT OD flows cannot be obtained in real-time because there is always trip duration time from the origin to the destination, as shown in Fig. 1. The OD matrix can only be obtained when all the travelers finish their trips. Therefore, when conducting real-time URT OD predictions, the model inputs should be carefully considered. One noteworthy point is that the model input for real-time road OD demand prediction is usually the road traffic volumes, which can be obtained in real-time (Xiong et al., 2019). The real-time ride-hailing OD demand can also be obtained in real-time as users need to identify the origin and destination when they start to use the services (Ke *et al.*, 2019). In addition, for the URT data, the inflow/outflow volume is always equal to the sum of OD flows as all passengers will eventually exit the stations.

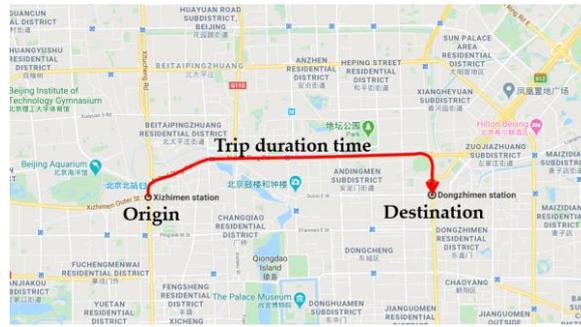

Fig. 1. Diagram of trip duration time of the OD pair

2. **Data dimensionality.** The number of OD flows is $n^2$ times the number of stations, where $n$ is the number of URT stations. This tremendously increases the difficulty in the prediction of OD flows. For example, there are 404 URT stations in Beijing, meaning the dimensionality of OD flow in Beijing is 163,216. Similarly, there are 270 stations in London and 424 stations in New York, and hence the dimensionality of OD flow is much larger than the cardinality of the transportation networks.

3. **Data sparsity.** The URT OD flow is spatiotemporally sparse, meaning that there is no passenger flow for many OD pairs. The reasons are two-fold: 1) the travel patterns change over time, leading to the time-varying sparse patterns of OD flows. For example, most OD flows depart from residual areas and arrive in the commercial areas in the morning peak hours, and the flow is reversed during the afternoon peak hours; 2) due to the large dimensionality of OD flow, the total travel demand disperses over different OD pairs, making many OD flows either small or zero. Fig. 2 presents the OD pairs from a downtown station (A) to suburb stations (B, C, D, E) in Beijing, and we can read from the data that there are generally small or zero OD flows between A→B, A→C, A→D, and A→E in the morning because few passengers go from downtown to suburb during morning peak hours. Table 1 shows the proportion of different OD flows in specific time intervals. There are more than 40% OD pairs with zero flow throughout the day. OD flows fewer than two account for more than 65% throughout the day. These small or zero OD flows are usually attributed to randomly generated trips that significantly decrease the regularity of OD flows, and thus increase the prediction difficulty. OD predictions for road traffic and ride-hailing service also suffers data sparsity issue while they are not as serious as that of URT, due to the availability of the real-time information. In contrast, the corresponding URT inflow/outflow, road flow, and ride-hailing demand are dense, and the volume is relatively large.

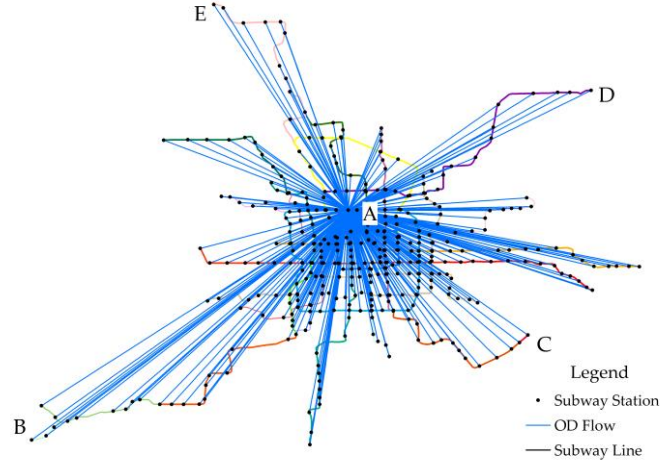

Fig. 2. Diagram of OD flows in Beijing, China

Table 1 OD flow statistics in OD matrix in Beijing, China

| OD flow | 05:00–05:30 | 08:00–08:30 | 12:00–12:30 | 19:00–19:30 |
|---------|-------------|-------------|-------------|-------------|
| OD = 0 | 95.16% | 40.53% | 58.46% | 54.92% |
| 0 < OD ≤ 2 | 4.34% | 25.21% | 27.57% | 25.41% |
| 2 < OD ≤ 4 | 0.34% | 10.40% | 7.89% | 8.22% |
| 4 < OD ≤ 6 | 0.09% | 5.74% | 2.87% | 3.77% |
| OD > 6 | 0.07% | 18.12% | 3.21% | 7.68% |

Table 2 summarizes the characteristics of different traffic state forecasting methods. As can be seen, the URT OD flow prediction task is the only task that requires careful design of model inputs, and the data dimensionality and sparsity issues also present. Given the valuable information in the URT OD flow, there is a lack of study on the short-term forecasting methods for the OD passenger flow for URT.

Table 2 Summary of the characteristics of different traffic state forecasting methods

|  | Data Availability | Data Dimensionality | Data Sparsity | References |
|--|-------------------|---------------------|---------------|------------|
| Road Speed/Flow | ✓ | ✓ | ✓ | (Guo *et al.*, 2019, Ma *et al.*, 2015) |
| Road OD Flow | ✓ | ! | ! | (Xiong et al., 2019) |
| Ride-hailing Demand | ✓ | ✓ | ✓ | (Geng *et al.*, 2019) |
| Ride-hailing OD Demand | ✓ | ! | ! | (Ke et al., 2019) |
| URT Inflow/outflow | ✓ | ✓ | ✓ | (Zhang *et al.*, 2020) |
| URT OD Flow | ! | ! | ! | This study |

Note: "✓" denotes issue not present and "!" denotes issue present.

In summary, this study is motivated by several issues to be addressed in short-term OD prediction in URT.

1. First, real-time OD matrices are unavailable. It is impractical to use OD matrices in the last several time intervals as model inputs. Hence, determining the inputs is the first problem to be tackled when conducting real-time OD prediction.

2. Second, existing studies generally ignore the relationship between inflow/outflow and OD flows in URT. Thus, their relationship should be explicitly modeled in the prediction models.

3. Third, most studies treat OD pairs with large flows and small flows equally, which can significantly reduce the prediction accuracy as there are many OD pairs with even no flow. The data sparsity issue is critical for OD flow prediction. Therefore, how to address the data sparsity to improve the prediction accuracy is another crucial problem.

4. Finally, in general, state-of-the-art deep-learning models are becoming increasingly complicated to improve prediction accuracy. However, the important question is whether increased complexity is better. This is another issue that needs to be explored.

In view of this, this paper introduces a channel-wise attentive split convolutional neural network (CAS-CNN) model to address these problems. In the proposed model, an inflow/outflow-gated mechanism is originally introduced to aggregate historical OD flow information and real-time inflow/outflow information, which solves the data availability issue. A split CNN is proposed for the first time in short-term OD prediction to combine sparse OD data and convert to dense features. Moreover, a masked loss function is introduced and justified mathematically to address the data dimensionality and sparsity issues. A channel-wise attention mechanism is applied to score the inputs as well as the extracted high-level features. Experiments on two real-world datasets from the Beijing subway show the superiority of the CAS–CNN model. The main contributions are summarized as follows:

1. The characteristics of URT OD prediction and the comparisons with other traffic forecasting tasks are summarized in detail. The problems of the short-term OD prediction in URT are also summarized.

2. An inflow/outflow-gated mechanism is developed to aggregate historical OD flow information and real-time inflow/outflow information by considering their intrinsic dependency.

3. A split CNN model is introduced to convert the sparse OD flow information to dense and useful features. To the best of our knowledge, this is the first time that the split CNN is introduced in short-term OD predictions.

4. A masked loss function is proposed based on the OD attraction degree (ODAD) indicator to handle small or zero OD flows.

The remaining sections are organized as follows. Section 2 reviews the literature. The methodology is described in Section 3. The experimental details and results are presented in Section 4, and the conclusions are summarized in Section 5.

## 2 Literature review

## 2.1 Traffic inflow and outflow predictions

Research studies on traffic inflow and outflow predictions have been prevailing in recent years, and the adopted methods range from conventional statistical methods to artificial-intelligence-based methods. The latter has been proved to be more effective in real-world applications owing to the massive mobility data that has been collected in recent decades, as well as the emerging deep-learning techniques. Since the long short-term memory (LSTM) was first introduced in the traffic prediction field in 2015 (Ma et al., 2015), many deep-learning models have been proposed, such as the classical CNN (Ma *et al.*, 2017), stacked autoencoder (Lv *et al.*, 2015), ST-ResNet (Zhang *et al.*, 2017), and ST-GCN (Yu *et al.*, 2017), as well as the latest hybrid models that combined two or more RNNs, CNNs, and GCNs, such as ResLSTM (Zhang et al., 2020), RSTN (Guo and Zhang, 2020), SBU-LSTM (Cui *et al.*, 2020), Conv-GCN (Zhang *et al.*, 2020b), TGC-LSTM (Cui *et al.*, 2019), GATCN (Guo and Yuan, 2020), and GA-LSTM (Zhang and Guo, 2020). Recently, the transformer (Xu *et al.*, 2020), the generative adversarial (Zhang *et al.*, 2019), and the capsule networks (Ma *et al.*, 2020) are also utilized for short-term predictions. Among these models, some are used for short-term predictions whereas others are for medium- or long-term predictions (Li *et al.*, 2018, Sun and Chen, 2019). Some are for single or several subway stations (Liu et al., 2019, Zhang et al., 2019), whereas others are for network-wide predictions. Some are for predictions under normal conditions (Xu et al., 2020, Jin *et al.*, 2020), whereas others are for abnormal conditions (Yu *et al.*, 2020). Some are for predictions using stationary correlations (Chai *et al.*, 2018), whereas others are for predictions using dynamic correlations (Yao *et al.*, 2019).

Overall, many types of models are built to accommodate various scenarios. However, all of them are for inflow or outflow predictions and are critically different from OD prediction in terms of data availability, data dimensionality, and data sparsity as mentioned in the introduction section. Therefore, it is necessary to build forecasting models that explicitly consider the unique characteristics of URT OD flows.

## 2.2 Traffic OD prediction and estimation

Due to the data dimensionality and sparsity issues, obtaining accurate OD demand is much more challenging than obtaining the inflow or outflow, regardless of whether the OD demand is road demand, ride-hailing demand, or URT demand.

Traffic OD prediction is different from inflow or outflow predictions, as mentioned in the introduction section. In terms of the prediction methods and the research objects, we have divided related studies into several categories as follows.

In terms of the methods, OD matrix prediction and estimation can be categorized into three categories. The first category is the conventional methods, such as the least-squares estimation algorithm (Yao *et al.*, 2016) and probability analysis model (Wang *et al.*, 2011). The second category is the machine learning method, such as the state space model (Yao *et al.*, 2015, Lin and Chang, 2007), back-propagation neural network (Zhou *et al.*, 2016), principal component analysis and singular value decomposition (Yang *et al.*, 2017), and hierarchical Bayesian networks (Ma *et al.*, 2013). However, there are some common shortcomings among these two categories. First, these methods cannot meet the real-time requirements. For example, when applied to large-scale networks, the least-squares method and state space model consume considerable computational resources, making them practically inapplicable. Second, the prediction accuracy needs to be improved. Third, the spatial and temporal correlations of the OD demand can hardly be considered.

To tackle these problems, deep-learning methods, which belong to the third category, have been developed extensively in recent years. Some researchers used the LSTM model to conduct the OD matrix prediction (Xi *et al.*, 2018). Each node is trained to obtain a specific LSTM model with the use of the parallel computing technique. However, this method cannot capture spatial correlations among all the OD pairs. Some studies applied CNN and GCN (Liu *et al.*, 2019, Wang *et al.*, 2019) to perform OD matrix prediction. These studies were applied to road traffic paradigms, in which the origin and destination zones are significantly different from the subway systems. Some studies (Xiong et al., 2019) also leveraged GCN to perform OD matrix prediction in road traffic paradigms in which links were treated as nodes and the adjacent matrix represented link connections. Destination prediction in the bike-sharing system was also explored by combining LSTM and CNN (Jiang *et al.*, 2019), while fewer bike stations could significantly reduce the problem difficulty comparing to the URT system. Overall, the contextual information of the above mentioned deep-learning studies is different from that of URT, while they can still provide intuition and implications for developing the deep learning model for URT OD flow prediction.

In terms of research objects, short-term OD matrix prediction or estimation can be divided into road OD estimation (Lin and Chang, 2007), taxi OD matrix prediction (Ou *et al.*, 2019, Liu et al., 2019, Wang et al., 2019), bus OD matrix prediction (Zhang *et al.*, 2017), and URT OD matrix prediction (Yang et al., 2017, Yao et al., 2016, Yao et al., 2015, Wang et al., 2011, Zhao *et al.*, 2007). The data available is different for different objects**.** In the road network, neither the real-time nor the true OD matrices cannot be obtained. However, the sectional link counts can be observed. Thus, the road OD matrix can be estimated via optimization models such as the bi-level programming model. Notably, it is difficult to evaluate the reliability of the estimated OD matrix because there is no true OD matrix for comparison (Yang *et al.*, 1991). In the case of the taxi OD matrix, because there are no fixed boarding and alighting points, existing methods always partition the entire research area to construct origin and destination regions (Traffic Analysis Zones or TAZs). In this case, the true OD matrix between TAZs can be obtained, whereas the real-time counterparts cannot. In the bus system, existing studies focus on one or several lines to conduct OD matrix prediction because the bus network is critically large-scale. Furthermore, data availability issue varies in different bus systems, as some bus systems can record boarding and alighting stations, whereas some can only record the boarding station. In URT, there are fixed subway stations, and passengers must swipe cards when they enter and exit stations. Therefore, the true OD matrix can be obtained based on historical smart card data. However, as discussed in the previous section, the real-time counterparts cannot be obtained because of the trip duration time.

Specifically, in URT, the OD matrix prediction studies that use deep-learning methods are critically few. Several existing studies built state space models (Chen *et al.*, 2017, Yao et al., 2015) or least-squares methods (Yao et al., 2016). Notably, a recent study applied the LSTM to perform OD matrix prediction in which a specific LSTM model was specifically trained for each of all the subway stations leveraging parallel computing techniques (Zhang et al., 2019). However, these studies exhibit some drawbacks. For example, they cannot capture complicated spatiotemporal correlations and nonlinear characteristics among OD flows. Moreover, each URT station requires to train a deep learning model separately, making the method computationally infeasible because there are many subway stations.

## 2.3 Summary

In order to contribute to the literature of short-term OD demand prediction, two main issues should be addressed for the URT system. Overall, the principle is that unique models should be developed because of the unique characteristics of the URT system. The detailed discussions are as follows:

(1) Deep learning methods have certain advantages over conventional methods. Many studies have demonstrated that deep learning methods can meet the real-time requirements as well as have high prediction accuracy. Moreover, it is feasible to train only one model for all stations. Therefore, developing deep-learning models to conduct OD prediction in URT is in real need.

(2) Due to the data availability issue, two types of information are available: OD demand in previous days and the real-time inflow/outflow. Hence it is critically important to combine the information of both data.

(3) The OD matrices are critically sparse and the dimension is large, especially in a large subway network. A systemic way needs to be developed to account for the sparsity level of different OD pairs and large dimensions of OD matrices.

# 3 Methodology

In this section, we formulate the methodological architecture. First, the problem of short-term OD prediction in URT is defined, and an indicator called the ODAD is introduced. The model architecture is then developed, followed by the introduction of the split CNN, the channel-wise attention mechanism, and the inflow/outflow-gated mechanism.

## 3.1 Problem definition

The goal of this study is to predict the OD matrix in the next time interval using historical information. The time interval is defined as 30 min in this study. The OD matrix $M$ and inflow/outflow $N$ can be extracted from smart card data in URT, and can be defined according to Eqs. (1) to (3). Notably, the OD flow of each time interval depends on the time interval in which the passengers enter the stations, as the exit time of each passenger might differ. The inflow/outflow series is extracted according to the corresponding entry station, entry time, exit station, and exit time.

$$M^{d,t} \in \mathbf{R}^{n \times n} = \begin{pmatrix} m_{11}^{d,t} & m_{12}^{d,t} & m_{13}^{d,t} & \cdots & m_{1j}^{d,t} \\ m_{21}^{d,t} & m_{22}^{d,t} & m_{23}^{d,t} & \cdots & m_{2j}^{d,t} \\ m_{31}^{d,t} & m_{32}^{d,t} & m_{33}^{d,t} & \cdots & m_{3j}^{d,t} \\ \vdots & \vdots & \vdots & \ddots & \vdots \\ m_{i1}^{d,t} & m_{i2}^{d,t} & m_{i3}^{d,t} & \cdots & m_{ij}^{d,t} \end{pmatrix} \quad (1)$$

$$N^{d,t} \in \mathbf{R}^{n \times t} = \begin{pmatrix} n_1^{d,1} & n_1^{d,2} & n_1^{d,3} & \cdots & n_1^{d,t} \\ n_2^{d,1} & n_2^{d,2} & n_2^{d,3} & \cdots & n_2^{d,t} \\ n_3^{d,1} & n_3^{d,2} & n_3^{d,3} & \cdots & n_3^{d,t} \\ \vdots & \vdots & \vdots & \ddots & \vdots \\ n_i^{d,1} & n_i^{d,2} & n_i^{d,3} & \cdots & n_i^{d,t} \end{pmatrix} \quad (2)$$

$$\begin{pmatrix} n_1^{d,t} \\ n_2^{d,t} \\ n_3^{d,t} \\ \vdots \\ n_i^{d,t} \end{pmatrix} = \begin{pmatrix} \sum_j m_{1j}^{d,t} \\ \sum_j m_{2j}^{d,t} \\ \sum_j m_{3j}^{d,t} \\ \vdots \\ \sum_j m_{ij}^{d,t} \end{pmatrix}, \quad n_i^{d,t} = \sum_j m_{ij}^{d,t} \ (infllow), \ or$$

$$\begin{pmatrix} n_1^{d,t} \\ n_2^{d,t} \\ n_3^{d,t} \\ \vdots \\ n_i^{d,t} \end{pmatrix} = \begin{pmatrix} \sum_i m_{i1}^{d,t} \\ \sum_i m_{i2}^{d,t} \\ \sum_i m_{i3}^{d,t} \\ \vdots \\ \sum_i m_{ij}^{d,t} \end{pmatrix}, \quad n_i^{d,t} = \sum_i m_{ij}^{d,t} \ (outflow)$$

$(3)$

where $m_{ij}^{d,t}$ represents the OD flow from station $i$ to station $j$ in the time interval $t$ on day $d$. $\sum_j m_{ij}^{d,t}$ is the sum of OD flows that enter station $i$ in the time interval $t$ on day $d$, and $n_i^{d,t}$ is the inflow/outflow entering/exiting station $i$ in the time interval $t$ on day $d$. The stations are ordered according to their adjacency in the subway line. Notably, there are inherent correlations between inflow/outflow and OD flows, as shown in Eq. (3). The inflow/outflow equals to the sum of corresponding OD flows in each row/column.

Regarding short-term OD prediction, prior studies generally used the OD matrices in the last several time intervals as model inputs to predict the OD matrix in the subsequent time interval (Liu et al., 2019, Wang et al., 2019). However, the real-time OD matrix cannot be obtained because of the trip duration time. Therefore, these studies cannot be applied for real-time operations. Similarly, in the real-time operation of URT, the real-time OD matrix cannot be obtained. However, real-time inflow/outflow is available. Therefore, this study seeks to predict the short-term OD matrix using the OD matrix of the previous several days, as well as the inflow/outflow of the same day, as expressed by Eq. (4).

$$M^{d,t} = f(\{M^{d-x,t}\}_x, \{N^{d,t-y}\}_y), \ x = 1,2,3\cdots; y = 1,2,3\cdots \tag{4}$$

where $M^{d,t}$ is the OD matrix in the time interval $t$ on day $d$. One of the inputs is the OD matrix $M^{d-x,t}$ in the same time interval $t$ during the last several days $d$-$x$. Another input is the inflow/outflow series $N^{d,t-y}$ during the last several time intervals $t$-$y$ of the same day $d$. Because the real-time OD matrix is not available, we innovatively designed an inflow/outflow -gated mechanism with the real-time inflow as inputs to provide real-time information.

## 3.2 Origin–destination attraction degree (ODAD) level

To characterize OD flows with different volumes, we introduce a novel indicator called ODAD (Zhang et al., 2019). It is defined as the average OD flow in a specific time interval during a longer period, as indicated by Eq. (5).

$$\text{ODAD}^t = \begin{pmatrix} a_{11}^t & a_{12}^t & a_{13}^t & \cdots & a_{1j}^t \\ a_{21}^t & a_{22}^t & a_{23}^t & \cdots & a_{2j}^t \\ a_{31}^t & a_{32}^t & a_{33}^t & \cdots & a_{3j}^t \\ \vdots & \vdots & \vdots & \ddots & \vdots \\ a_{i1}^t & a_{i2}^t & a_{i3}^t & \cdots & a_{ij}^t \end{pmatrix}, a_{ij}^t = \frac{1}{n}\sum_{d=1}^{n} a_{ij}^{d,t} \tag{5}$$

where $a_{ij}^t$ is the average OD flow from station $i$ to station $j$ during an $n$-day period. It is a dynamic indicator that varies with time. For a specific OD pair, the $a_{ij}^t$ value may be low early in the morning and high during peak hours. It is also an average indicator used to avoid randomness.

To handle OD pairs with different attraction degrees, we divide all OD pairs into five levels according to their ODAD values, as shown in Table 3. The variation of OD numbers at different ODAD levels is shown in Fig. 3. A sub-OD matrix in a single time interval is shown in Fig. 4. Temporally, the OD pairs at low and lowest levels account for a large majority. Spatially, OD flows only occur in specific areas. These small values negatively affect the model performance because the lack of regularity increases the difficulty to make predictions. Therefore, it is challenging to handle these small or zero values. To solve this problem, we innovatively introduce a masked loss function according to the "low" ODAD level in Section 3.7, thus reducing the impact of small or zero OD flows on the prediction accuracy. The "low" ODAD level used in this study is fixed and does not change with time.

Table 3 OD attraction degree (ODAD) level definition

| ODAD value | $a_{ij}^t = 0$ | $0 < a_{ij}^t \leq 2$ | $2 < a_{ij}^t \leq 4$ | $4 < a_{ij}^t \leq 6$ | $a_{ij}^t > 6$ |
|---|---|---|---|---|---|
| ODAD level | Lowest | Low | Middle | High | Highest |

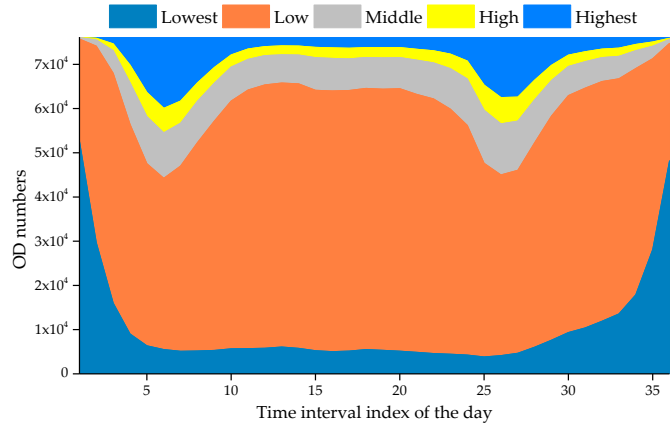

Fig. 3. Variation of OD numbers in a single day

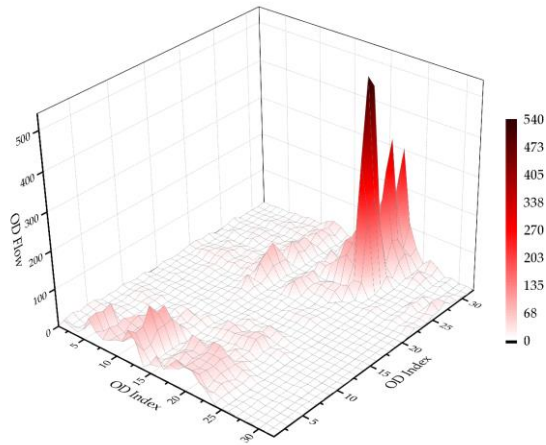

Fig. 4. Three-dimensional (3D) view of a sub-OD matrix

### 3.3 Model development

We propose the prediction framework based on the split CNN, channel-wise attention, and inflow/outflow -gated mechanism (referred to as CAS–CNN as shown in Fig. 5) to conduct short-term OD prediction in URT. The CAS–CNN comprises two branches for historical data and real-time data, respectively.

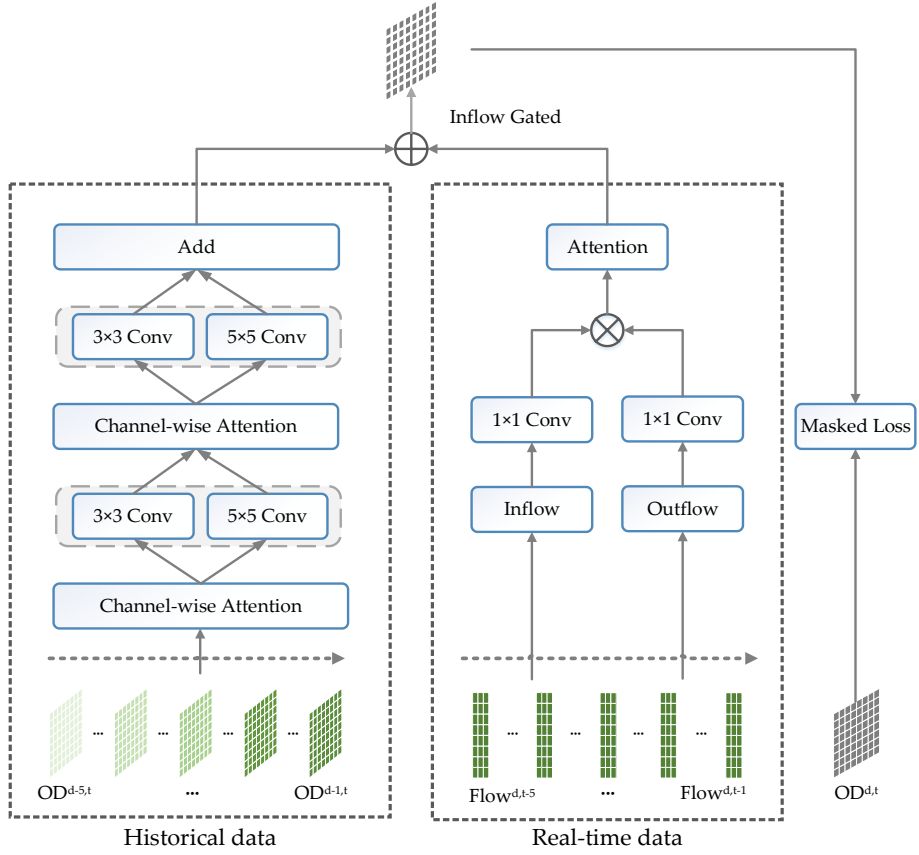

Fig. 5. Model architecture for CAS–CNN

In the branch of the historical data (short as trunk), we originally introduce a split CNN to capture spatiotemporal correlations with different perceptive fields, as well as to produce dense information from sparse OD flows. The channel-wise attention is used to weight the inputs, and different high-level features are extracted from the OD matrix. To the best of our knowledge, this is the first time that the split CNN is applied to URT OD prediction.

In the branch of real-time information, we use real-time inflows/outflows as inputs to extract important information. To merge the two sources of data, an ingenious inflow/outflow-gated mechanism is designed to aggregate historical OD flow information and real-time inflow/outflow information by considering their intrinsic dependency.

To address small and zero OD flows, we also introduce a masked loss function based on the low ODAD level.

In the following sections, the split CNN, channel-wise attention, inflow/outflow-gated mechanism, and the masked loss function are described in detail.

## 3.4 Split CNN

Existing studies generally use one same-size kernel to extract features (Ma et al., 2017, Zhang et al., 2020, Liu et al., 2019). In this case, to improve the training performance, a general method is to increase the network depth (number of layers). However, the increase in the number of layers has multiple undesirable effects, such as overfitting, vanishing gradient, gradient explosion, etc. Although the residual network (He *et al.*, 2015) has been proposed to solve these problems, it also increases network complexity and computational resources such as the training time.

Motivated by GoogLeNet (Szegedy *et al.*, 2015, Szegedy *et al.*, 2016), in this study, we originally introduce a split CNN model to address the task of short-term OD prediction. To the best of our knowledge, this is the first time that the split CNN is applied to short-term OD prediction in URT. Rather than deepen the network, we choose to widen it with different kernels, as shown in Fig. 6 and Fig. 7, as this can effectively increase the adaptability of the network.

As mentioned above, one of the issues in short-term OD prediction in URT is data sparsity. The split architecture is exactly suitable for OD matrices in URT because of the serious data sparsity problem in two aspects.

Temporally, as shown in Table 1 and Fig. 3, OD flows at the "lowest" ODAD level (namely, zero OD flows) are more than 40% throughout a day. By designing the split architecture, dense data can be generated from relatively sparse matrices, and more information can be extracted via different-size kernels. It cannot only increase the performance of neural networks but also ensure the training efficiency.

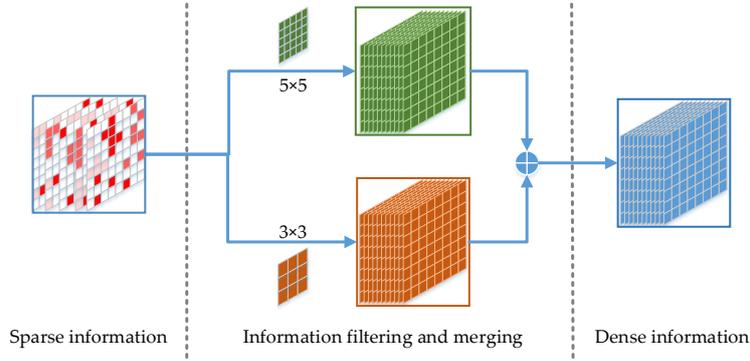

Fig. 6. Diagram of split CNN (From sparse information to dense information)

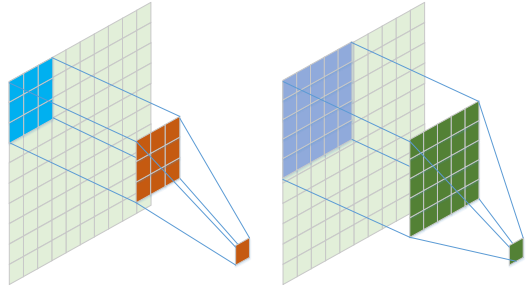

Fig. 7. Diagram of the 3×3 and 5×5 kernels

Spatially, as shown in Fig. 4, only some specific areas exist OD flows. Therefore, in flattened areas, a smaller kernel is adequate to capture its spatial characteristics. However, in peak areas, a larger kernel is more suitable because it can capture more information with the use of a larger perception field. In this case, some important information cannot be easily omitted.

To this end, we introduce a split CNN for OD prediction in URT. The values $v$ at position $(x, y)$ in the $j^{th}$ feature map of the $i^{th}$ layer can be calculated as follows (Zhang et al., 2020b).

$$v_{ij}^{xy} = \left( b_{ij} + \sum_m \sum_{p=0}^{P_s-1} \sum_{q=0}^{Q_s-1} w_{ijm}^{pq} v_{(i-1)m}^{(x+p)(y+q)} \right)_{k_1} + \left( b_{ij} + \sum_m \sum_{p=0}^{P_s-1} \sum_{q=0}^{Q_s-1} w_{ijm}^{pq} v_{(i-1)m}^{(x+p)(y+q)} \right)_{k_2} \qquad (6)$$

where $m$ denotes the index of the feature map in the $(i\text{-}1)^{th}$ layer, $w_{ijm}^{pq}$ is the $(p, q)^{th}$ value of the kernel connected to the $m^{th}$ feature map in the $(i\text{-}1)^{th}$ layer, $(P, Q)$ denotes the kernel dimension, and $k$ denotes kernels with different sizes.

### 3.5  Channel-wise attention

The human-visual attention mechanism is a type of brain signal processing mechanism of human vision. By quickly scanning the global image, human vision acquires the target area that needs attention. Then, more attention resources are devoted to this area to obtain more detailed information about the target. Other useless information was suppressed simultaneously. This is a mechanism used by humans to quickly select high-value information from a large amount of information with limited attention resources. The human visual attention mechanism significantly improves the efficiency and accuracy of visual information processing.

Motivated by human visual attention, many types of attention mechanisms, such as self-attention and position-wise attention in Transformer (Vaswani *et al.*, 2017), residual attention (Wang *et al.*, 2017), multilayer attention (Yang *et al.*, 2016), and spatial attention (Chen *et al.*, 2017) have been proposed. The channel-wise attention mechanism was first proposed by (Chen et al., 2017). It was used to weigh different high-level features, and can be applied in OD prediction for several aspects.

On the one hand, in the field of OD prediction in URT, the real-time OD matrix is not available. Therefore, we use the OD matrix in the same time interval of the last several days as one of the model inputs, as shown in Fig. 5. However, some of the OD matrices are highly related to the outputs. Some are lowly correlated to the outputs. It is taken for granted that the channel-wise attention can be used to weigh different OD inputs.

On the other hand, the output of split CNN represents high-level features extracted from inputs. It is important to adaptively focus more on some critical features to improve model performance. Therefore, we innovatively apply the channel-wise attention mechanism into the output of split CNN and add them together as shown in Fig. 8. Fig. 9. shows the details of channel-wise attention. There is a tensor reduction $R$ during tensor processing used to represent nonlinear features. The output can thus be expressed as follows.

$$O^l = \begin{pmatrix} Y^l = CNN(X^l) \\ \delta_1 = \Phi(Y^l) \\ O_1^l = Y^l \times \delta_1 \end{pmatrix}_{k_1} + \begin{pmatrix} Y^l = CNN(X^l) \\ \delta_2 = \Phi(Y^l) \\ O_2^l = Y^l \times \delta_2 \end{pmatrix}_{k_2} \qquad (7)$$

where $\Phi$ represents the channel-wise attention operation, $\delta$ is the attention vector, and $k$ denotes different kernels.

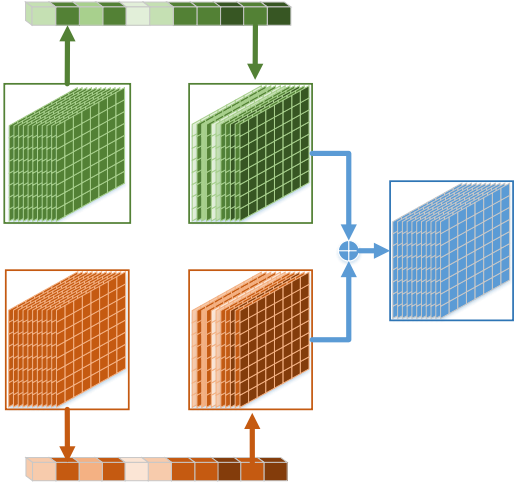
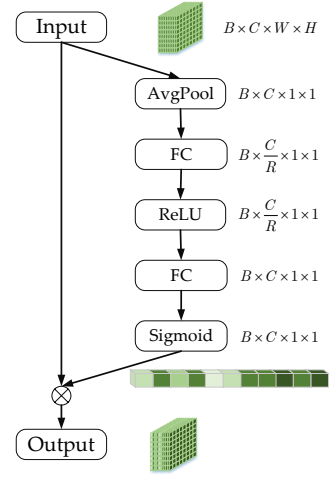

Fig. 8. Channel-wise attention after the split CNN          Fig. 9. Diagram of channel-wise attention

## 3.6  Inflow/outflow-gated mechanism

As is mentioned above, the real-time OD matrix is not available in URT. How to conduct OD predictions incorporating real-time information is dramatically important. As shown in Eqs. (1) to (3), there are strong correlations between inflows/outflows and OD flows. Motivated by the relationship, we originally introduce an inflow/outflow-gated mechanism to effectively control the trunk output and fuse the inflow/outflow and OD matrix information, as shown in Fig. 10. The inflow/outflow go through a 1×1 convolutional layer. Their outputs are multiplied and then are weighted by an attention parameter vector. The weighted inflow features plus the output of the model trunk according to the rows is followed by a 1×1 convolution to obtain the final predicted results.

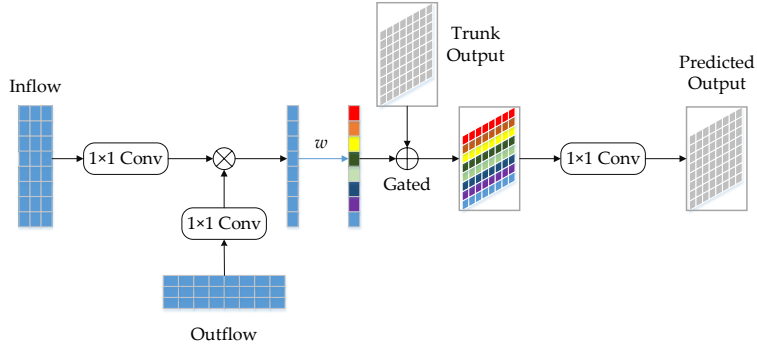

Fig. 10. Diagram of inflow/outflow-gated mechanism

The inflow/outflow $N^{d,t} \in \mathrm{R}^{n \times t}$ in the last several time intervals is processed as follows.

$$
\begin{aligned}
O_{inflow/outflow} &= w \times CNN_{1\times1}(N_{inflow}) \times CNN_{1\times1}(N_{outflow}) \\
O_{fuse} &= \left( O_{inflow/outflow} + O_{trunk} \right)_{by\ rows} \\
O_{pre} &= CNN_{1\times1}(O_{fuse})
\end{aligned}
\tag{8}
$$

where $w$ is a column vector denoting the attention parameters of $N_{inflow}$ and $N_{outflow}$, and $O_{trunk}$ denotes the output of the model trunk. The variable $w$ is used to model interpretability analysis, as detailed in Section 4.3.3.

Notably, a 1×1 convolutional layer is applied to obtain the final output. Each 1×1 convolutional kernel can realize cross-channel information communication. The 1×1 kernel can replace the fully connected layer when nonlinear features are captured, while model complexity is reduced. Therefore, although this denotes a simple linear combination, it is conducive to information fusion and feature extraction.

## 3.7 Masked loss function

As discussed in previous sections, there are numerous small or zero OD flows that significantly affect the prediction performance. Moreover, OD flows in different ODAD levels are highly imbalanced both temporally and spatially, as shown in Table 1, Fig. 3, and Fig. 4. Therefore, we introduce a masked loss function (M-Loss) as Eq. (9). We construct a mask file according to the low ODAD level to mask the OD flows whose ODAD level are less than two (Zhang et al., 2019).

$$M\text{-}Loss = MSE = \frac{1}{(n \times n)_{no\_mask}} \sum_{i,j} mask \times (m_{ij}^{d,t} \text{-} m_{ij}^{\overset{d,t}{}})^2 \qquad (9)$$

where $MSE$ is the masked mean-squared error, $(n \times n)_{no\_mask}$ indicates the OD numbers that are not masked, $m_{ij}^{d,t}$ is the actual value and is the same as that in Eq. (1), $m_{ij}^{d,t}$ is the corresponding predicted value, and the *mask* is a matrix file, with values of zeros and ones, which indicates whether the corresponding OD flows are masked or not. Because the ODAD value changes with time for a specific OD pair, the mask file will be correspondingly updated with time. It is noted that if the values are masked, the errors are not backpropagated here, as proved in Eqs. (10) to (12), which can significantly improve the prediction performance of OD flows that are not masked. The *mask* highlights the import OD pairs, with higher traffic volumes. Only the errors of important flows (i.e. OD flows with high volumes) will be backpropagated here. Assuming $y = w \times x$,

$$MSE = \frac{1}{(n \times n)_{no\_mask}} \sum_{i,j} mask \times (m_{ij}^{d,t} \text{-} w \times m_{ij}^{d,t})^2, \qquad (10)$$

$$gradient = \frac{\partial MSE}{\partial w} = \frac{-2 \times m_{ij}^{d,t}}{(n \times n)_{no\_mask}} \sum_{i,j} mask \times (m_{ij}^{d,t} \text{-} w \times m_{ij}^{d,t}), \qquad (11)$$

$$w_{new} = w - lr \times gradient = w + lr \times \frac{2 \times m_{ij}^{d,t}}{(n \times n)_{no\_mask}} \sum_{i,j} mask \times (m_{ij}^{d,t} \text{-} w \times m_{ij}^{d,t}). \qquad (12)$$

If the value $m_{ij}^{d,t}$ is masked, the corresponding *mask* is zero. Therefore, $w_{new} = w$, thus indicating that the errors are not backpropagated.

# 4 Experiment

In this section, we test the proposed method with two real-world datasets and compare it with benchmark methods. The experimental results are also analyzed from multiple perspectives.

## 4.1  Data description

Two datasets from the Beijing Subway are used in the experiments, as shown in Table 4. There are 276 and 308 stations in MetroBJ2016 and MetroBJ2018, respectively. We use 25 weekdays from consecutive five-week periods. The data records from MetroBJ2018 are fewer than those from MetroBJ2016 because more people entered the stations using the QR code on a mobile phone rather than swiping cards in 2018. Each record contains the card number, entry-station name, entry time, exit-station name, and the exit time. The OD matrix and inflow/outflow series can be extracted according to Eqs. (1) and (2) every 30 min. All data are normalized using the min-max scaler.

Table 4 Data description

| Description | MetroBJ2016 | MetroBJ2018 |
|---|---|---|
| Date | February 29, 2016 to April 3, 2016 | October 8, 2018 to November 11, 2018 |
| Time | 05:00 to 23:00 | 05:00 to 23:00 |
| Week number | 5 | 5 |
| Data record | 130 million | 110 million |
| Station number | 276 | 308 |
| Matrix dimension | 276 × 276 | 308 × 308 |
| Time interval | 30 min | 30 min |
| Matrix number in a day | 36 | 36 |

## 4.2  Model configurations

The data from the first four weeks are for training and validating the model, while the rest are for testing the model. The validation rate is set to 0.1. The early stopping technique is applied during model training to avoid overfitting. The training and validation losses are shown in Fig. 11. According to the parameter tuning results, as shown in Fig. 12, we determine the hyperparameters of time steps, filters, batch size, and R (reduction). For the inflow/outflow-gated branch, the inflow/outflow series in the last five time steps (2.5 h) in the entire network are utilized. There is one layer with 16 filters for the first split CNN and one layer with one filter for the second split CNN in the trunk. The learning rate is 0.001 and the batch size is 16. The tensor reduction $R$ is set to two in the channel-wise attention. We use the OD matrix in the same time interval in the last five days. We use the Xavier normal initializer to initialize the CNN related parameters. All models are implemented with PyTorch on a desktop computer with Intel i7-8700K processor (12M cache up to 3.20 GHz), 24 GB memory, and an NVIDIA GeForce GTX 1070Ti graphics card

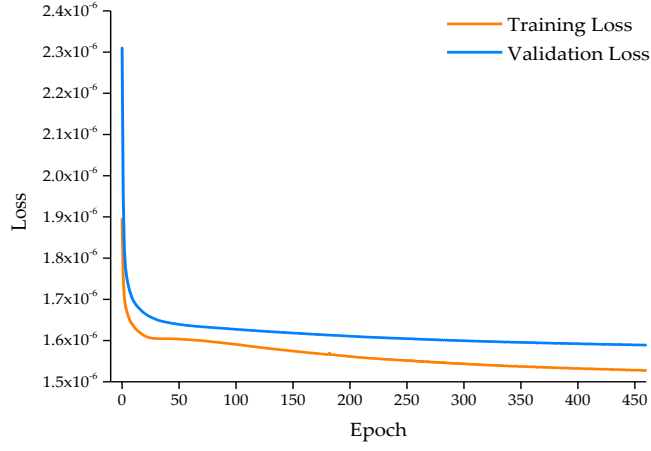

Fig. 11 Variation of training loss and validation loss

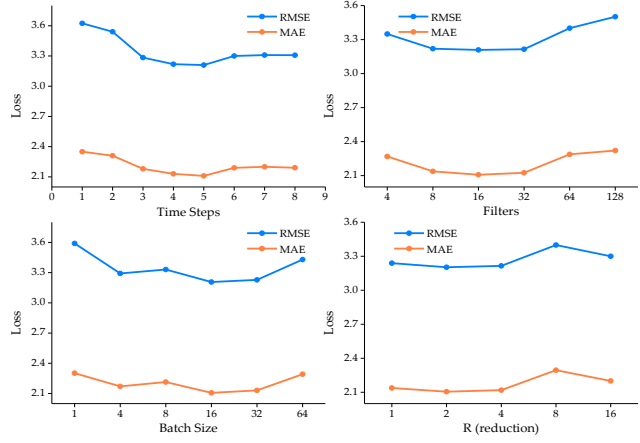

Fig. 12 Parameter tuning results

## 4.3 Evaluation metrics

In this study, the root-mean-squared error (RMSE), mean absolute error (MAE), and weighted mean-absolute-percentage error (WMAPE) are chosen as evaluation metrics according to Eqs. (13) to (15).

$$RMSE = \sqrt{\frac{1}{(n \times n)_{no\_mask}} \sum_{i,j} \left(m_{ij}^{d,t} - \overset{\wedge}{m}_{ij}^{d,t}\right)^2} \tag{13}$$

$$MAE = \frac{1}{(n \times n)_{no\_mask}} \sum_{i,j} \frac{\left|m_{ij}^{d,t} - \overset{\wedge}{m}_{ij}^{d,t}\right|}{m_{ij}^{d,t}} \tag{14}$$

$$WMAPE = \sum_{ij} \left( \frac{m_{ij}^{d,t}}{\sum_{ij} m_{ij}^{d,t}} \left| \frac{m_{ij}^{d,t} - \overset{\wedge}{m}_{ij}^{d,t}}{m_{ij}^{d,t}} \right| \right), \quad m_{ij}^{d,t} \rangle 0 \tag{15}$$

The notations are the same as those in Eq. (9). The $(n \times n)_{no\_mask}$ exactly indicates the OD numbers that are not masked.

## 4.4 Benchmarks

In this section, we compare the proposed CAS–CNN model with several other models,

including 2D CNN, 3D CNN, ConvLSTM, ConvGRU, TrajGRU, and ST-ResNet. Moreover, we built another five models based on our model to prove the effectiveness of the proposed split CNN, the masked loss function, the channel-wise attention mechanism, and the inflow/outflow-gated mechanism. For all of them, all other parameters except the control component are the same as those in CAS–CNN. For CAS–CNN, we construct a mask file based on the low ODAD level, and apply it to the M-Loss function. The inputs and outputs are the same for all models. Note that we do not include the classical models like the historical average or autoregressive integrated moving average models in the baseline modes. Because they are unable to make predictions for a metrics (with more than 76,176 OD pairs) using only one model. The detailed information for each benchmark is listed as follows.

**2D CNN and 3D CNN:** Both of them have three layers with 8, 16, and 1 filters, respectively. The activation function for the first two layers is ReLU and the last layer is linear. The kernel size is 5×5. The learning rate is 0.001. The batch size is 8.

**ConvLSTM** (Xingjian *et al.*, 2015) **and ConvGRU** (Shi *et al.*, 2017)**:** Both of them have three layers with 8, 8, and 1 filters, respectively. The kernel size is 3×3. The learning rate is 0.001. The batch size is 8.

**TrajGRU** (Shi et al., 2017)**:** This is an encoder–forecaster architecture. There are two layers with 32 and 64 filters, respectively, in the encoder part. There are two layers with 64 and 32 filters, respectively, in the forecaster part. The respective kernel sizes are 3×3 and 5×5. The learning rate is 0.0001. The batch size is 16.

**ST-**ResNet (Zhang *et al.*, 2017): We use the residual block similar as one branch of the original ST-ResNet.

**CAS–CNN (No S-CNN):** We replace the split CNN with general CNN to prove the effectiveness of the split CNN. The kernel size is 3×3 in the general CNN.

**CAS–CNN (No Mask):** We replace the M-Loss with the general MSE loss function to prove the effectiveness of the M-Loss.

**CAS–CNN (No CA):** We delete the channel-wise attention mechanism to prove the effectiveness of the mechanism.

**CAS–CNN (No Inflow):** We delete the inflow-gated branch to prove the effectiveness of the inflow-gated mechanism.

**CAS–CNN (No Outflow):** We delete the outflow-gated branch to prove the effectiveness of the outflow-gated mechanism.

**CAS-CNN:** The whole model we propose in section 3.3.

## 4.5  Results and discussions

### 4.5.1 Network-wide prediction performance

The network-wide prediction performance is shown in Table 5 and Fig. 13. Several critical points can be drawn as follows:

1.  It is not the more complicated the better for the deep learning models. It is the more appropriate the better. For the benchmarks, 2D CNN and 3D CNN perform similarly, and with the same regularity as those of ConvLSTM and ConvGRU. However, notably, the simpler CNNs perform better than ConvLSTM and ConvGRU. Three reasons may account for this finding. First, the ConvLSTM or ConvGRU are proposed for precipitation nowcasting whose input data are denser than OD matrices in URT, especially for the matrices early in the morning or very late in the night. Second, ConvLSTM and ConvGRU are more suitable for series that have a strict chronological order. However, because the real-time OD matrices cannot be obtained in URT, the input data for this study is obtained from five former days. The chronological order is not obvious. Third, we only have consecutive five-week AFC data. When a relatively complicated method is trained, more data is necessary. Therefore, a simpler model might perform relatively better when the data volume is limited. Moreover, the TrajGRU, which is the most complex baseline model has the worst performance. This is because the TrajGRU is mainly designed for location-variant motion patterns that may not be suitable for OD matrices in URT. These results indicate that increased complexity might be not beneficial for the deep-learning models. Increased appropriateness is more beneficial.

2. The proposed inflow/outflow-gated mechanism is conducive to improve the prediction performance. This indicates that the use of inflow/outflow series to replace the real-time OD matrices and to provide real-time information is effective. From the CAS–CNN (No Inflow/Outflow) to CAS–CNN, the RMSE is improved for MetroBJ2016 and MetroBJ2018 by 0.84% and by 1.90%, respectively. These values denote average improvements for an individual OD flow. From the network point-of-view, it is a significant improvement because there are many OD pairs in a single time interval and there are ten million passengers taking the subway in one day in Beijing, China.

3. The proposed split CNN, the masked loss function, and the channel-wise attention mechanism are also proved to be effective for model performance. This shows that deliberately dealing with small or zero flows contributes to the improvement of model performance. From the CAS–CNN (No S-CNN) to CAS–CNN, the RMSE is improved for MetroBJ2016 and MetroBJ2018 by 1.60% and by 3.16%, respectively. From the CAS–CNN (No Mask) to CAS–CNN, the RMSE is improved for MetroBJ2016 and MetroBJ2018 by 0.93% and by 1.96%, respectively. From the CAS–CNN (No CA) to CAS–CNN, the RMSE is improved for MetroBJ2016 and MetroBJ2018 by 0.87% and by 1.86%, respectively. Irrespective of the case, the CAS–CNN performs the best, which benefits from the architecture of split CNN, channel-wise attention mechanism, inflow/outflow-gated mechanism, and the masked loss function.

Table 5 Comparison of performances of different models

| Models | MetroBJ2016 | | | MetroBJ2018 | | |
|---|---|---|---|---|---|---|
| | RMSE | MAE | WMAPE | RMSE | MAE | WMAPE |
| 2D CNN | 3.266 | 2.154 | 26.94% | 3.185 | 1.986 | 27.54% |
| 3D CNN | 3.271 | 2.157 | 27.00% | 3.193 | 1.993 | 27.66% |
| ConvLSTM | 3.284 | 2.157 | 26.98% | 3.109 | 2.008 | 27.83% |
| ConvGRU | 3.307 | 2.168 | 27.10% | 3.121 | 2.015 | 27.86% |
| TrajGRU | 3.643 | 2.349 | 29.46% | 3.800 | 2.457 | 34.56% |
| ST-ResNet | 3.260 | 2.149 | 26.81% | 3.179 | 1.971 | 27.49% |
| *CAS–CNN (No S-CNN)* | *3.255* | *2.142* | *26.68%* | *3.099* | *1.991* | *27.42%* |
| *CAS–CNN (No Mask)* | *3.233* | *2.130* | *26.58%* | *3.061* | *1.973* | *27.30%* |
| *CAS-CNN (No CA)* | *3.231* | *2.121* | *26.40%* | *3.058* | *1.963* | *27.00%* |
| *CAS–CNN (No Inflow)* | *3.230* | *2.127* | *26.38%* | *3.059* | *1.962* | *26.98%* |
| *CAS–CNN (No Outflow)* | *3.229* | *2.117* | *26.37%* | *3.057* | *1.958* | *26.79%* |
| **CAS–CNN** | **3.203** | **2.105** | **26.10%** | **3.001** | **1.905** | **26.33%** |

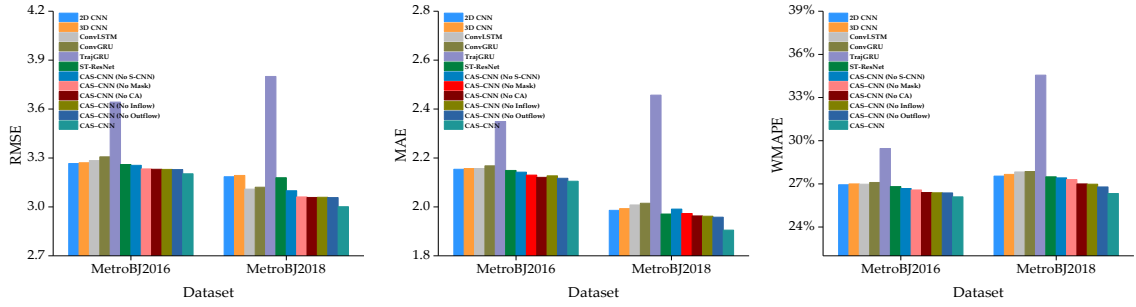

Fig. 13 Comparison of performances of different models

### 4.5.2 Prediction performances of individual OD pairs

To evaluate models on individual OD flows, we choose several OD flows to compare the actual values and predicted values, as shown in Fig. 14 and Fig. 15. As is shown, OD_1 is an OD flow with peak features in the morning hours. No matter for MetroBJ2016 or MetroBJ2018, the CAS–CNN can accurately capture the variation throughout a day. Even the peak flows can be predicted accurately. For OD_2 and OD_4, one is an OD flow with peak features in the morning hours, and the other with peak features in the evening hours. Both of them exhibit significant variations throughout the day. However, it can be observed that the trend can be captured even under the case of significant variations. For OD_3, the flows undergone large volume reductions from 2016 to 2018, resulting in some flows in off-peak hours are masked. As is shown in Fig. 15, even when the flows in some time intervals are masked, the CAS–CNN model performs well throughout the day. In summary, the proposed CAS–CNN can perform well on an individual level in most cases in two real-world subway datasets.

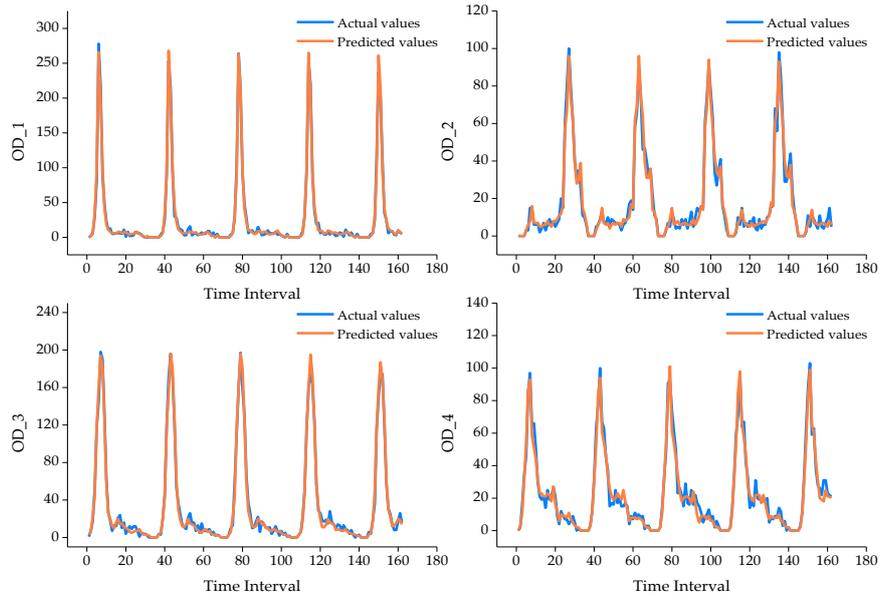

Fig. 14 Comparison of actual and predicted flows of four randomly selected OD pairs in MetroBJ2016

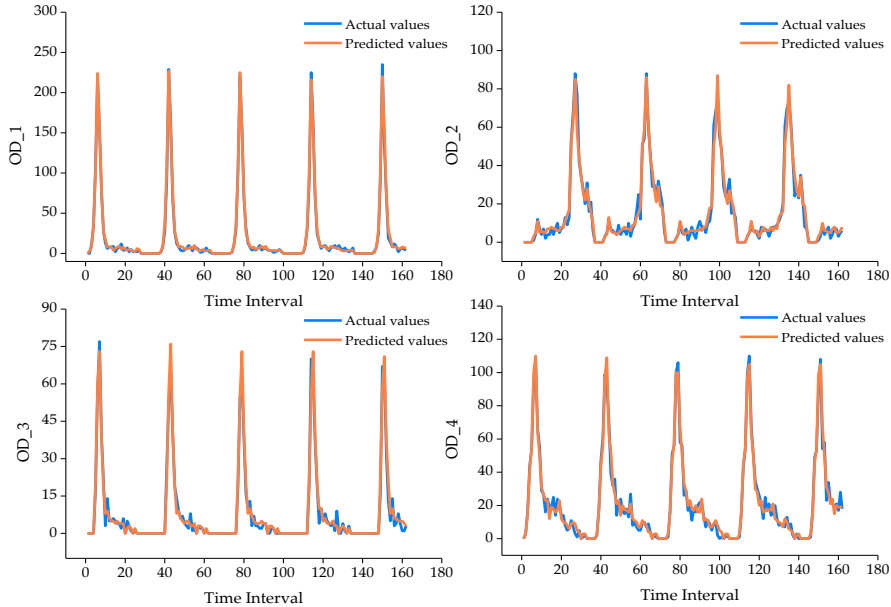

Fig. 15 Comparison of actual and predicted flows of four randomly selected OD pairs in MetroBJ2018

### 4.5.3 Prediction performance in different time intervals

To evaluate the models in different time intervals, we calculate the average prediction accuracy at each time interval. The trend of the average evaluation metrics with time is shown in Fig. 16. In particular, there are flow vibrations at 02:00 pm, thus leading to the change of the prediction accuracies for all models. Therefore, we also compare the model performance during this period. The enlarged line graph from Fig. 16 is shown in Fig. 17. Several findings are listed as follows.

1. The performance for different models in different time intervals presents the same patterns as the overall performance. The CAS-CNN outperforms the rest whether in peak hours or off-peak hours. The results demonstrate the stability of the CAS-CNN model.

2. Models perform stably in both morning and mid-night. When it comes to peak hours, the performance gap becomes large, indicating that the CAS-CNN can capture the ridership variation better.

3. There is flow variation from 01:30 pm to 03:00 pm, leading to the performance variation for all models. However, the CAS-CNN performs the best by a large margin during this period as shown in Fig. 17, indicating its strong adaptability.

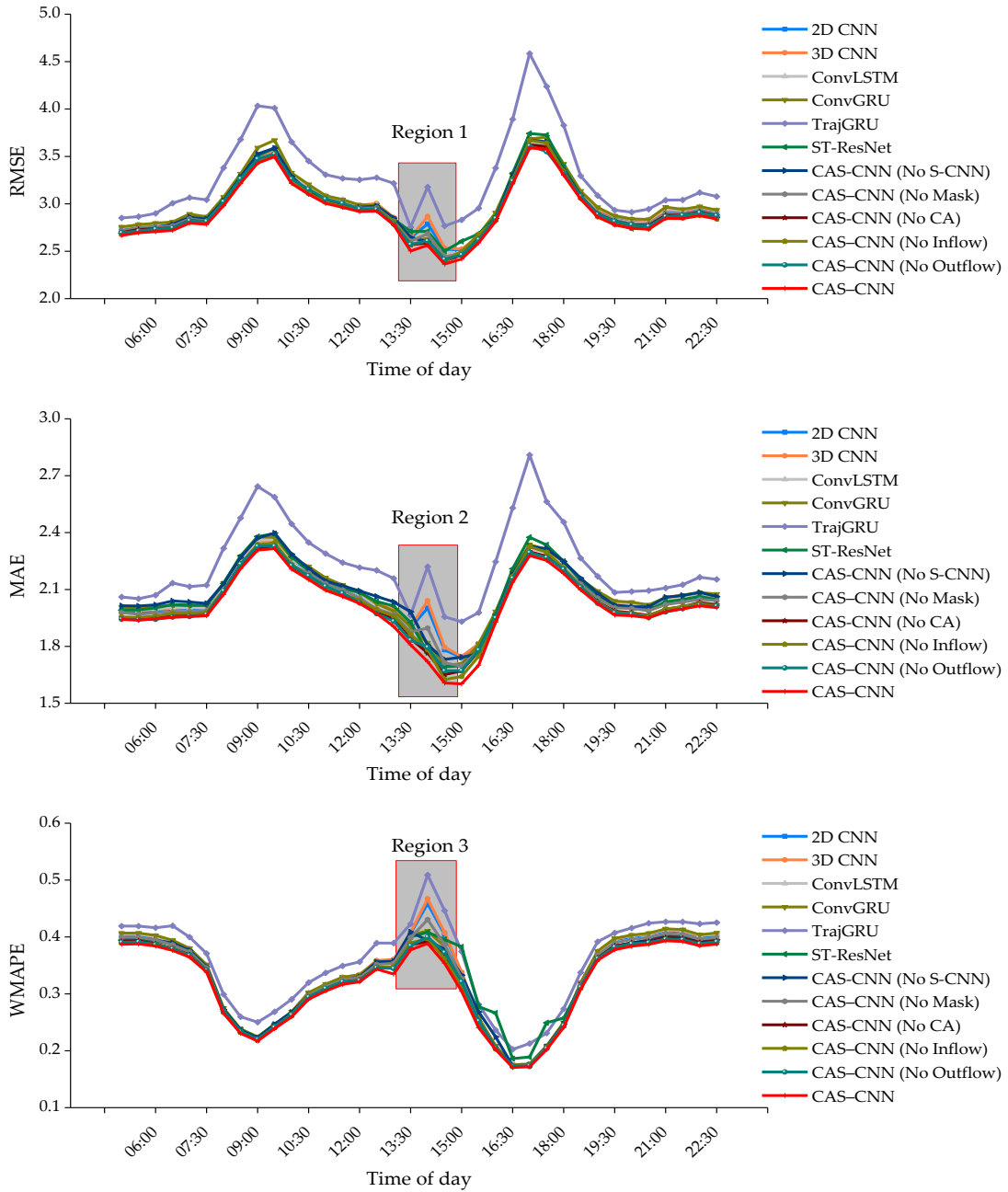

Fig. 16 Comparison of the model performance in different time intervals

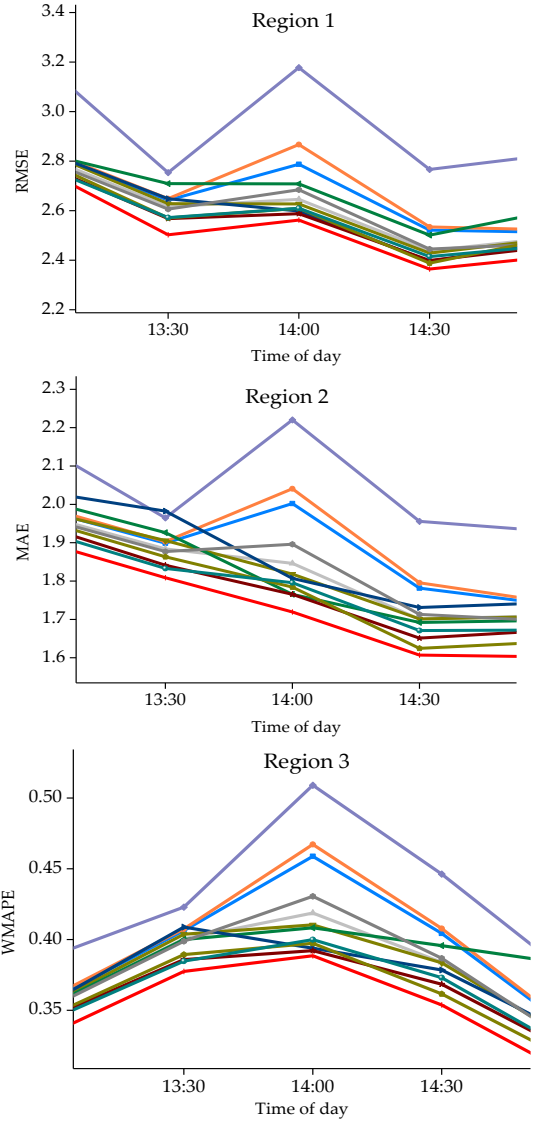

Fig. 17 Enlarged subregions from Fig. 15

### 4.5.4 Model interpretability

Because the real-time OD matrices are unavailable in URT, we introduce the inflow/outflow-gated mechanism to provide real-time information. To further explore the effect and the meaning of the inflow/outflow-gated mechanism, as well as the influence of small OD flows in the model, we have plotted the relationship between the inflow volume and $w$ in Eq. (8), namely, the gated parameters, as shown in Fig. 18. For convenience, we choose 100 stations and the corresponding parameters to display. The parameters are the final trained ones. The inflow volume is the sum of inflows from the corresponding subway station. They are normalized using the min-max scaler.

As is clearly shown, there is an obvious negative correlation between the two series. This also proves the effect and the meaning of the proposed inflow-gated mechanism. Two reasons can account for this. First, the inflow-gated mechanism is mainly designed to control the output of the trunk. Therefore, the parameter size represents the strength of the control. Second, small OD flows always produce large errors. To reduce the errors, the branch needs to adjust the small flows as more as possible. However, for large OD flows, it is unnecessary to make large adjustments. Therefore, the larger the inflow volume is, the smaller the corresponding parameter is.

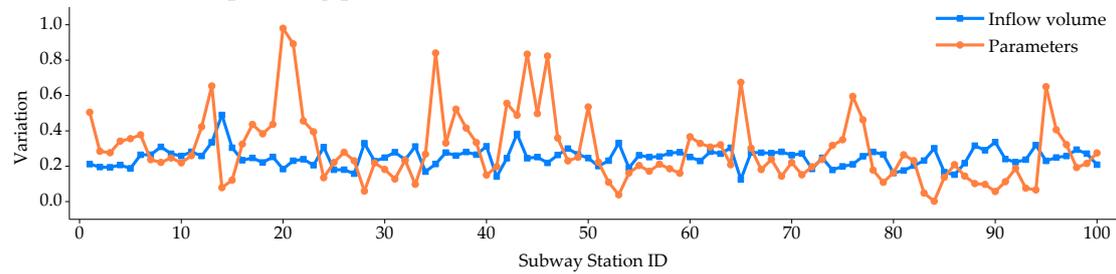

Fig. 18 Relationship between the inflow volume and the gated parameters

# 5 Conclusions

This study proposes a channel-wise attentive split convolutional neural network (CAS–CNN) model to conduct short-term OD prediction in URT. The proposed model consists of various novel components, such as the split CNN, the channel-wise attention mechanism, inflow/outflow-gated mechanism, and the masked loss function to address the unique issues that lie in the URT OD prediction. In particular, the proposed model is able to address the serious data sparsity issue with the use of a user-specified mask. The split CNN is also able to obtain dense information from sparse OD matrices. To the best of our knowledge, this is the first time that the split CNN is applied to short-term OD prediction in URT. Given that the real-time OD matrices in URT are unavailable, we innovatively introduce an inflow/outflow-gated mechanism to merge the historical OD demand information with real-time inflow/outflow information. The main findings of the study are summarized as follows:

1. Deep-learning models are becoming increasingly complex recently. Results show that models are not the more complicated the better. It is the more appropriate the better.

2. The data sparsity and data dimensionality issues of OD flow are critical problems that need to be solved in URT. The proposed split CNN is able to produce dense information from sparse OD flows. The proposed masked loss function is proved to be effective in improving the prediction accuracy when the OD flow under low ODAD level is masked.

3. Real-time OD flow information is unavailable because of the trip duration time. The originally introduced inflow/outflow-gated mechanism can process the real-time inflow/outflow information and merge them with the historical OD information, and it improves model performance. In addition, the developed gated mechanism demonstrates good model interpretability.

4. The CAS-CNN model demonstrates strong stability and adaptability in different time intervals, especially under the case of flow variations.

Overall, these findings can provide critical insights for real-time subway operation and management. In the future, multi-source data (weather conditions, road congestion, accidents) can be used to further improve the prediction accuracy. How to determine the threshold values of the ODAD levels is also an issue to be explored. Time information such as time of the day and day of the week can also be considered to improve the performance. Besides, the OD flow prediction issue on the weekends needs to be addressed owing to the different travel patterns. More methods can be explored to address the data sparsity issue and the lack of real-time information in short-term OD prediction in the URT system.

# Conflicts of Interest

The authors declare no conflict of interest.

# Acknowledgments

We wish to thank the anonymous reviewers for the valuable comments, suggestions, and discussions. This work was supported by the National Natural Science Foundation of China (Project No. 71871010 and 71871027), and the grant funded by the Hong Kong Polytechnic University (Project No. P0033933) .